# ViQIE: A New Approach for Visual Query Interpretation and Extraction

Radhouane Boughammoura
Department of Computer Sciences
Faculty of Sciences of Monastir,
Research Unit MARS
Monastir, Tunisia
Radhouane.Boughammoura@gmail.com

Lobna Hlaoua
Department of Electronic and Computer Science
High School of Sciences and Technology of H.Sousse,
Unité de Recherche MARS
H. Sousse, Tunisia
Lobna1511@yahoo.fr

Mohamed Nazih Omri
Department of Computer Sciences
Faculty of Sciences of Monastir,
Research Unit MARS
Monastir, Tunisia
MohamedNazih.Omri@fsm.rnu.tn

*Abstract*— **Web services are accessed via query interfaces which hide databases containing thousands of relevant information. User's side, distant database is a black box which accepts query and returns results, there is no way to access database schema which reflect data and query meanings. Hence, web services are very autonomous. Users view this autonomy as a major drawback because they need often to combine query capabilities of many web services at the same time. In this work, we will present a new approach which allows users to benefit of query capabilities of many web services while respecting autonomy of each service. This solution is a new contribution in Information Retrieval research axe and has proven good performances on two standard datasets.**

I. INTRODUCTION

Nowdays, Web services offer query interfaces which give access to data located in distant databases. Interfaces are rendered in Web browser as web pages. They are interpreted visually by users in order to understand the meaning of query. So, he is able tofill correctly fields of the interface ans submit the query to database and obtains relevant results (see figure 1).

In User's side, a web service is a black box which has as input query and as output results from database. Query is described by the interface, it offers some query capabilities. For example, query interface of figure1 offer the capability to search roundtrip flights from departure city to destination city, some search criteria are added to the query such as departure date, return date, and number of passengers.

Query interfaces are designed in order to be perceived by human users, while we aims to make query interfaces accessible to machines. In client side, a query interface is a HTML script. Some approaches try to extract the query from this script. We have remarked that visual query and HTML query are very different because designers of query interface use visual editors to design interfaces and the HTML script is generated automatically. Two fields are visually close to each other in visual interface (such as departure city and destination city), but they are far to each other in HTML script because of query condition instances and HTML tags.

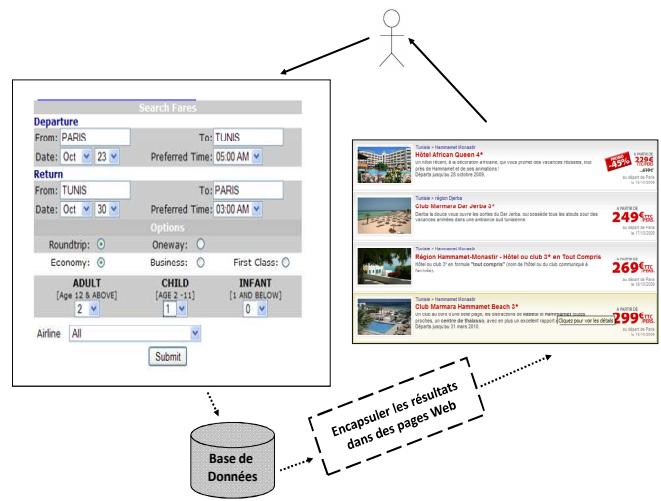

**Figure 1. Research on the Web Process**

We aims to bring the capacity of users to interpret visual query interfaces to machines. This will facilitate information retrieval from deep web databases. Using our approach we can send automatically queries to distant databases and obtaining relevant results while respecting usually the autonomy of the web service.

In this paper we present a new approach for query extraction from visual query interface. Section 2 gives our query model. Section 3 is our approach to extract visual query interface. We evaluate performance of our approach in section 4. Finally we conculdre and present our pespectives in section 5.

Our goal is to facilitate the task of searching for information to the user and this by offering a new generic interface that brings together the research capabilities of several Web services both (see figure 3). L'autonomie de chaque

## I. RELATED WORK

In literature, many researchers have become interested in modeling of the motion. We have noted the existence of two types of modelling: the flat model of the motion and the hierarchical model.

Madhaven et al [1] have represented the motion by a flat representation based on the textual form of the query. The text representation of the query is one where the values of the fields are set in the URL of the Web service. For example: ''http://jobs.com/find?src=hp&kw=chef&st=Any&sort=salary&s=go'' is a query to search for offers of used by keyword (kw) and governorate (st). This representation is a static representation of the query. This type of query is indexed by search engines.

Wensheng et al [2] have represented the motion by a hierarchical representation that describes the geometrical arrangement of the objects in the query in the IHM (see figure 4). In this case the query is Visual and is not indexed by search engines. However, the query is dynamic and varies depending on the data entered by the user.

Since the purpose of our work is to build the generic interface, we chose the Visual form of the motion, i.e. the hierarchical representation.

## I. MODEL OF REPRESENTATION OF THE QUERY

To facilitate the processing of the request by machine, we have defined a representation of the query that identifies the different concepts which composed and semantic relations that exist between them.

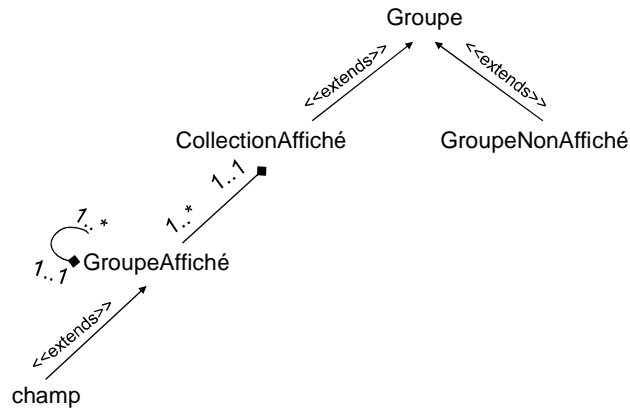

Figure 4. Representation of the query template

A search interface is a Visual representation of the query. Our model describes the motion on the basis of its visual representation: a query is an interweaving of several concepts that form the query. A concept (**GroupeAffiché**) is described in an interface by a group of fields (**field**). A group of concepts form a collection (**CollectionAffiché**). Some concepts are not displayed in the page (**GroupeNonAffiché** )).

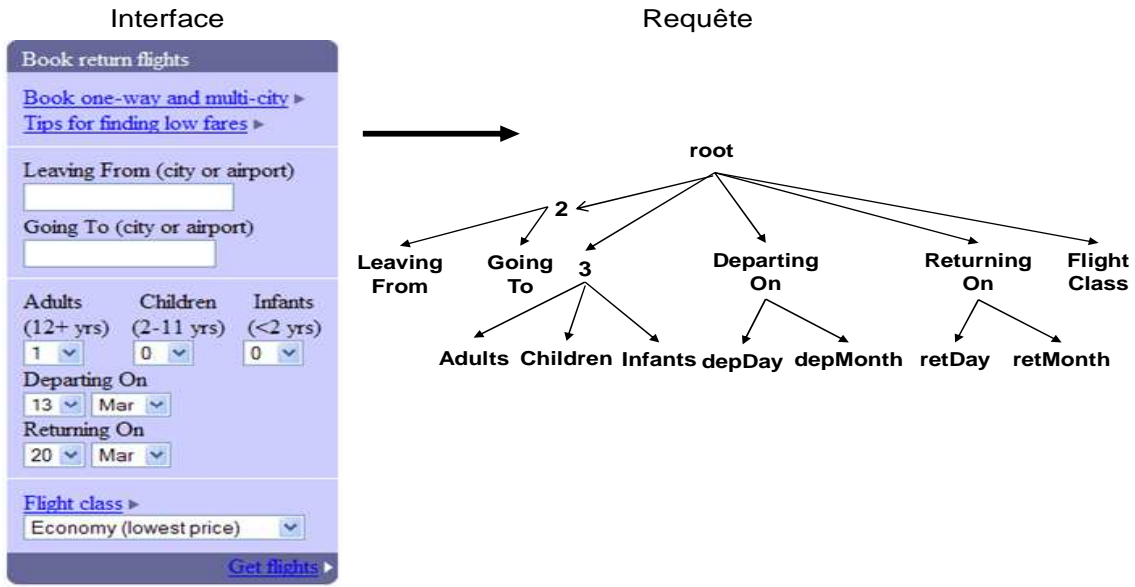

**Figure 2. Hierarchical representation of the query**

## II. EXTRACTION OF MOTION FROM THE INTERFACE

In a first step, it is to identify the query that will be submitted to the Web service. This service is often available through a search interface for information encoded by a HTML script and containing the different fields of the query in addition to the extra-informations of style (e.g. HTML tags). These information shall specify how fields in the Web page appariassent. In our contribution, we try to extract the query from the Web page. That is, identify the different objects which are the motion and semantic relations connecting them from the HMI. Figure 4 shows in the left part the HMI and in the right part of the query. In the structure of a query, we distinguish the different fields of the interface and the groups of fields. For example, the number of passengers (object "3" in figure 4) is a group of three fields 'Adults', 'Children', and 'Infants'. There is a semantic relationship "is - a" between the number of passengers object and son fields for each field is a passenger.

**Definition** *(Motion): A motion is a tree of items ordered. Each leaf is a field of the interface. Each inner element has an ordered set s (|s| > 1) of subelements, each can be a leaf or even internal node. The subelements are ordained by the sequence of their fields match (If sheet) or even groups of items (if internal node) in the interface.*

We have noticed that an object of the motion is often represented by a group of adjacent and aligned fields. For example in the interface of the figure, the fields 'Adults', 'Children', 'Infants' are adjacent and are on the same line. They are all aligned low alignment. These two Visual descriptors (adjacency and alignment) are very relevant indices for the collection of groups of fields.

We measured the adjacency between the fields by the Euclidean distance between two fields. This distance is defined by the Euclidean distance between two points p1 and p2 closest belonging respectively to the two fields f1 and f2. It is defined by the following formula:

$$DistEucl(f1, f2) = min_{p1 \in f1, p2 \in f2} DistEucl(p1, p2) \quad (1)$$

To measure the alignment between the fields, we used four types of alignment: alignment bottom, top, right, left. AlignX function returns 1 if two fields are aligned following X alignment:

$$AlignX(f1, f2) = \begin{cases} 1 \text{ si } f1 \text{ et } f2 \text{ sont alignés } X \\ 0 \text{ sinon} \end{cases} \quad (2)$$

With X = bottom (B), high (H), right (D), or even left (G)

The value of the alignment between two fields f1 and f2 takes into consideration the four types of alignment. The coefficient of low alignment is two times higher than other alignments coefficients because fields are on the same line (the bottom aligned) have a strong chance of belonging to a group.

$$Align(f1, f2) = AlignG(f1, f2) + AlignD(f1, f2) + AlignB(f1, f2) + 2 \cdot AlignB(f1, f2)$$

$$(3)$$

The distance between two fields f1 and f2 [2] is a combination of the Euclidean distance and the alignment between the two fields. Over this distance is small (respect. high), over the probability of membership of the two fields to the same group is high (low compliance.).

$$Distance(f1, f2) = \frac{DistEucl(f1, f2)}{Align(f1, f2)} \quad (4)$$

To determine the groups of fields in the interface, we used an algorithm of clustering by density DBSCAN (Density-Based Spatial Clustering of Applications with Noise) [3]. The definition of a cluster in DBSCAN is based on the notion of scope of density: a field f1 is within reach of density of a field f2 if f2 is a less remote to ε F1. We set ε at the minimum distance between any pair of non-grouped fields. Figure 5 is an illustration of this concept, we applied it on the interface in figure 4. Each black square is a field of the interface. We note that the fields corresponding to the "number of passengers" are density range because the scope of each field (green circle) reached the field the following in the group. Therefore the three fields "Adults", "Children", and "Infants" belong to a same cluster.

We note that the fields corresponding to the "number of passengers" are density range because the scope of each field (green circle) reached the field the following in the group.

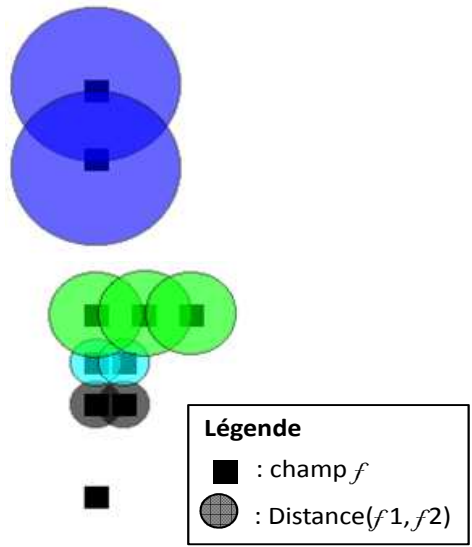

**Figure 3. Determination of groups of fields**

We have modified the algorithm DBSCAN in a way that it applies recursively on the fields and groups of fields. Thus obtained and the sweat-groups of fields. The judgment of our algorithm condition is reached when all the fields in the interface are grouped, it gets in this case the cluster "root" (see figure 4) that represents the query.

### III. EXPERIMENTAL RESULTS

ViQIE were assessed on the basis of standard test ICQ [4]. ICQ is a collection of 100 search interfaces of information and their queries manually extracted. These interfaces belong to 5 areas of interest: flights air, automotive, books, apartments, and offers of jobs.

Our method of evaluation is to extract the query from the interface and then compare it with the query manually extracted. If the two motions are told that our algorithm correctly extracts the query, if it is said that our algorithm has committed a fault. Table 1 summarizes the results.

We measured the accuracy of our approach to extraction of motions on the two bases of test ICQ and TEL-8. Figures 7 and 8 present the details of our algorithm on two test databases.

**Table1. Our approach for extracting query results**

|  | TEL-8 | | | ICQ | | |
|---|---|---|---|---|---|---|
|  | Flights | Cars | Books | Vols | Cars | Books |
| total | 20 | 19 | 19 | 20 | 20 | 19 |
| #correct queries | 13 | 14 | 17 | 13 | 13 | 16 |
| #errors | 7 | 5 | 2 | 7 | 7 | 3 |
| Precision | 0,65 | 0,73 | 0,89 | 0,65 | 0,65 | 0,84 |

We note that the accuracy of our approach depends on the complexity of the query: the more fields is high over the accuracy of our approach is low. And more level of nesting of the objects in the query, the more the accuracy of our approach is low. We also noted that the more complex queries are observed in the field of air flights and simple queries are observed in the field books.

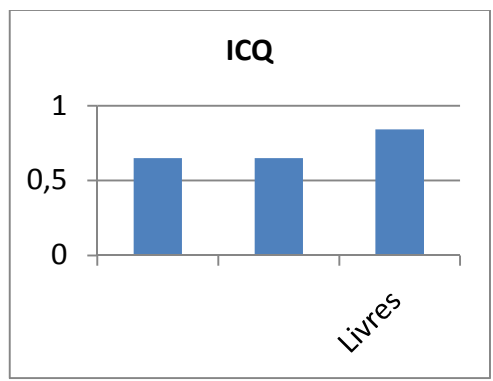

**Figure 4. Accuracy of our algorithm of extraction based on test ICQ**

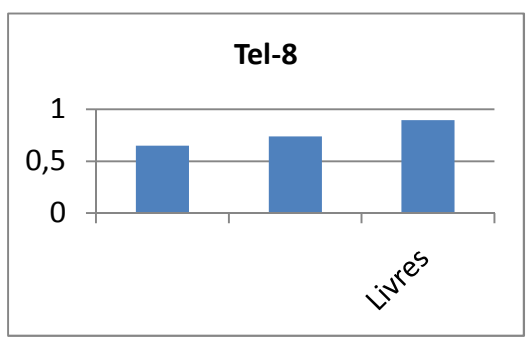

**Figure 5. Accuracy of our algorithm on the basis of test TEL-8**

**Conclusion**

In this paper we have presented two contributions: the first is a new technique of modeling concepts of motion and semantic relations linking it. The second contribution is the establishment of a new approach to extraction of motions based on visual interpretation of the interface and allows the identification of a query from existing interface. We experienced our approach based on two previous contributions and we found that on the two bases of standard tests ICQ and such-8 the performance of our approach are rather good.

Our future work is to propose a new approach for the integration of motions. This approach will allow the establishment of a new Web service which will facilitate the search for information in the invisible Web.

REFERENCES

[1] J. Madhavan, D. Ko, L. Kot, V. Ganapathy, A. Rasmussen, A.Y. Halevy. Google's Deep Web crawl. In Proceedings of VLDB, 2008.

[2] Wensheng, W., Doan, A-H, Yu, C., Meng, W ., Modeling and Extracting Deep-Web Query Interfaces, In Proceedings of AIIS 2009, 2009.

[3] Ester M., Kriegel H-P, Sander J., Xu X(1996-), *A density-based algorithm for discovering clusters in large spatial databases with noise*, *Proceedings of the Second International Conference on Knowledge Discovery and Data Mining (KDD-96), 1996.*

[4] *The UIUC Web Integration Repository,* Computer Science Department, University of Illinois at Urbana-Champaign, *http://metaquerier.cs.uiuc.edu/repository, 2003*

[5] *Gionis A., Mannila H., Tsaparas P. Clustering aggregation.* ACM Trans. Knowl. Discov. Data *1, 1, Article 0.4, 2007.*